# High-energy electronic excitations in La$_3$Ni$_2$O$_7$ by time-resolved optical spectroscopy

Junzhi Zhu[1,2,3#], Mengwu Huo[4#], Yubin Wang[5], Yuxin Zhai[5], Lili Hu[1], Haiyun Huang[1,2,3], Xiu Zhang[1,2,3], Baixu Xiang[5], Mengdi Zhang[1], Yusong Gan[5], Zhiyuan An[1,2,3], Meng Wang[4*], Qihua Xiong[1,5,6,7*], Haiyun Liu[1*]

[1] Beijing Academy of Quantum Information Sciences, Beijing 100193, China;
[2] Beijing National Laboratory for Condensed Matter Physics, Institute of Physics, Chinese Academy of Sciences, Beijing 100190, China;
[3] School of Physical Sciences, University of Chinese Academy of Sciences, Beijing 100049, China;
[4] Center for Neutron Science and Technology at School of Physics, Guangdong Provincial Key Laboratory of Magnetoelectric Physics and Devices, Sun Yat-Sen University, Guangzhou, 510275, Guangdong, China;
[5] State Key Laboratory of Low-Dimensional Quantum Physics and Department of Physics, Tsinghua University, Beijing 100084, China;
[6] Frontier Science Center for Quantum Information, Beijing 100084, China;
[7] Collaborative Innovation Center of Quantum Matter, Beijing 100084, China.
# These authors contributed equally to this work.
*To whom correspondence should be addressed.
*Emails:* M. W. wangmeng5@mail.sysu.edu.cn, Q.X. qihua_xiong@tsinghua.edu.cn and H.L. liuhy@baqis.ac.cn

**ABSTRACT**

Recently, high-temperature superconductivity has been established in bilayer La$_3$Ni$_2$O$_7$, which exhibits a density-wave (DW) transition at ~ 150 K under ambient pressure. The DW order is believed to be linked to superconductivity, as it is suppressed upon the emergence of superconductivity at high pressures. Here, we explore the ultrafast dynamics of high-energy electronic excitations from 10 K to room temperature under ambient pressure using time-resolved optical spectroscopy. Two high-energy electronic excitations at ~1.8 and ~ 2.4 eV, arising from distinct interband transitions, are identified. They exhibit different DW gaps of approximately 54 and 67 meV, respectively, along with relaxation dynamics that can be well described by the Rothwarf-Taylor model. In addition, we observe four coherent Raman-active phonon modes that exhibit distinct coupling with different electronic excitations. The phonon softening with increasing temperature can be well described between ~100 K and room temperature by a semi-quantitative model, which includes thermal expansion and anharmonic phonon-phonon coupling. At cryogenic temperatures, deviations from the measured temperature-dependent phonon frequencies and the model fits suggest an additional contribution from electron-phonon coupling. Our study provides direct evidence of the complex gap structure and phonon dynamics in this material, offering critical insights into the DW mechanism and many-body effects.

# I. INTRODUCTION

The recent discovery of superconducting nickelate compounds has established a new platform for investigating high-temperature superconductivity[1]. A representative nickelate superconductor is the Ruddlesden-Popper-type bilayer $La_3Ni_2O_7$, which exhibits a superconducting transition temperature of ~ 80 K in bulk form under high pressure[2-7] and ~ 40 K in thin films at ambient pressure[8,9]. Superconductivity appears in bulk $La_3Ni_2O_7$ under high pressure upon the suppression of a competing density-wave (DW) order characterized by a transition of $T_{DW}$ ~ 150 K at ambient conditions[4,10-16]. Notably, further studies have suggested that the DW order in these materials is not a single phase but rather involves a spin-density-wave (SDW) transition at 140 -150 K[11,14,15,17] and a charge-density-wave (CDW) transition at ~110 K[18]. From the electronic band perspective, $La_3Ni_2O_7$ is proposed to be either a Mott insulator[19], a system proximate to Mottness[18], or a charge-transfer type insulator [20,21], with correlated and multiple band structures[20,22-26]. Understanding the DW and electron dynamics is essential to elucidating the superconductivity mechanism.

Intriguingly, time-resolved optical spectroscopy based on femtosecond (fs) pump-probe pulses enables investigation of the ultrafast physical processes of strongly correlated materials unattainable at equilibrium, including the gap dynamics and the time evolution of interactions among electrons, phonons, spins, and other degrees of freedom[27]. In this experimental method, a pump pulse generates a population of hot carriers. The subsequent ultrafast relaxation of these hot carriers, mediated by scattering processes (*e.g.*, electron-electron and electron-phonon scattering), is then monitored by a time-delayed probe pulse. For materials with DW or superconducting gaps, the relaxation step exhibits a characteristic bottleneck effect, wherein hot carriers temporally accumulate above the gaps[28]. The boson-mediated over-gap recovery results in temperature-dependent gap dynamics that can be phenomenologically described by the Rothwarf-Taylor (RT) model[29]. Time-resolved optical spectroscopy has been well established as a powerful tool for unveiling gap dynamics in CDW[30-35], SDW[36,37], and cuprate superconducting materials[28,37-40]. Notably, compared a monochromatic/narrowband probe, a broadband white-light-continuum (WLC) probe provides an additional energy/frequency degree of freedom, enabling access to a comprehensive investigation of the high-energy electronic excitations[41], band renormalization[42,43], and coherent phonon modes[44] in cuprates. Recent time-resolved optical studies of $La_3Ni_2O_7$ have primarily employed monochromatic/narrowband lights (with a center wavelength of ~ 800 nm or ~ 400 nm) as pump and probe to investigate temperature-dependent DW dynamics at ambient conditions[45,46] and the evolution of DW gap at high pressure[47]. Despite these progresses, comprehensive investigations of the ultrafast dynamics of high-energy electronic excitations, their connection to the DW order, and coherent phonon dynamics, remain largely unexplored.

In this work, we present ultrafast dynamics of the high-energy electronic excitations from time-resolved optical spectroscopy with a WLC broadband probe. Our results reveal two distinct high-energy electronic excitations, both of which exhibit robust red shifts upon cooling below the DW transition, signaling the opening of two DW gaps.

The two DW-gap structure is further confirmed by the temperature-dependent over-gap relaxation, which are well fitted by the RT model. In addition, we observe four coherent phonon modes, with three showing a clear softening upon increasing temperature. The temperature-dependent phonon frequencies present deviate from the semi-quantitative anharmonic phonon model that includes both thermal expansion and phonon-phonon anharmonicity at cryogenic temperatures, suggesting a contribution of electron-phonon coupling in this regime.

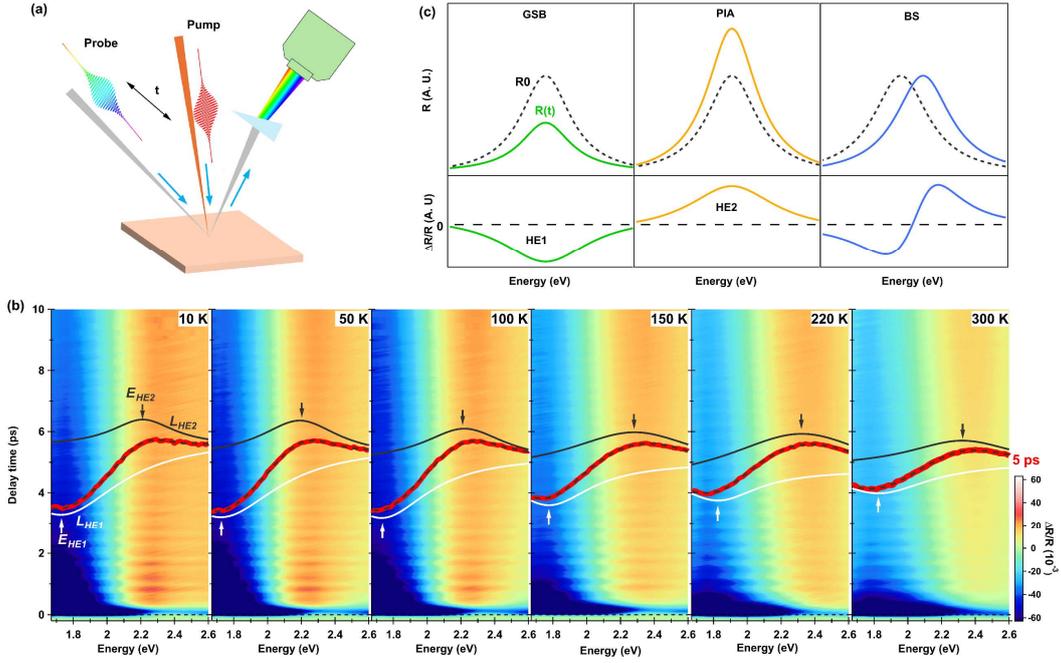

FIG. 1. Temperature-dependent high-energy electronic excitations in $La_3Ni_2O_7$. (a) Illustration of time-resolved optical spectroscopy. The sample was excited by pump pulses at the fluence of ~ 40 μJ/cm². The ultrafast dynamics were detected by WLC probe pulses in terms of transient reflection changes ΔR/R. (b) ΔR/R images as a function of photon energy and delay time. The red curves are horizontal cuts at 5 ps. The black dashed curves are fits consisting of a Lorentzian dip ($L_{HE1}$, white solid curves), a Lorentzian peak ($L_{HE2}$, black solid curves) and a constant background. The white and black arrows mark the dip and peak position ($E_{HE1}$ and $E_{HE2}$), respectively. (c) Mechanism of ground state bleaching (GSB), photo-induced absorption (PIA) and blue shift (BS). R0 and R(t) curves represent reflectivity without and with pump excitation, respectively. GSB (PIA) induces negative (positive) ΔR/R, corresponding to $HE1$ ($HE2$) signals. BS results in negative (lower energy side) and positive (higher energy side) ΔR/R signals.

## II. RESULTS
### A. IDENTIFICATION OF HIGH-ENERGY ELECTRONIC EXCITATIONS

In our setup, a femtosecond laser beam (center wavelength 800 nm, pulse duration 35 fs, and repetition rate 1 kHz) was divided into two paths by a beam splitter. One part served as the pump beam to excite the sample, while the other was focused into a sapphire crystal to generate the WLC broadband probe *via* nonlinear effects. Both pump and probe beams were focused onto the sample surface at near-normal incidence. The pump-probe delay was controlled by a motorized linear stage. Immediately following the pump excitation, the probe monitors the photoinduced transient reflectivity changes

ΔR/R, which were recorded by a spectrometer (Fig. 1a). Further details regarding the experimental setup are provided in Supplemental Material (Section 1) and can also be found in our previous publications[48,49].

Figure 1b shows ΔR/R images as a function of probe photon energy and pump-probe delay, from 10 to 300 K. At room temperature, the ΔR/R spectral curves exhibit a negative dip at ~ 1.8 eV and a positive peak at ~ 2.4 eV. The energy positions of these two features lie within the energy range where the optical properties are characterized by interband transitions[18,20,50]. This transient dip-peak structure cannot be simply explained by band renormalization (*i.e.*, a blue shift of an interband transition, as illustrated in Fig. 1c), because the dip and peak signals exhibit distinct kinetics and coherent oscillations (as shown in Figs. 2-4), which will be discussed in detail later. Instead, the observed ΔR/R spectra are better described as two high-energy electronic excitations (*HE1* and *HE2*), arising from independent ground-state bleaching (GSB) and photo-induced absorption (PIA) of distinct interband transitions. GSB manifested as a negative ΔR/R signal arises from the electron filling of conduction bands and the depletion of ground state electrons, and blocks further absorption of the probe[43,49,51]. On the contrary, PIA yields a positive ΔR/R signal, where the probe is more effectively absorbed by photoexcited carriers in nonequilibrium states[43,49,51], as illustrated in Fig. 2c.

Following the above analysis, we successfully modeled the ΔR/R spectra by a combination of a negative dip and a positive peak ΔR/R(E) = $L_{HE1}$ + $L_{HE2}$ + C, where $L_{HE1}$ and $L_{HE2}$ are the Lorentzian dip and peak at $E_{HE1}$ = 1.83 eV and $E_{HE1}$ = 2.33 eV, respectively. The constant term C accounts for the high-energy extension of the Drude component, analogous to the Drude contribution as observed in cuprate superconductors[41]. Comparison with the DFT calculations by B. Geisler *et al*. in Ref. [50], allows us to assign the lower-energy feature $L_{HE1}$ to the in-plane transition between Ni $3d_{x^2-y^2}$ and Ni $3d_z^2$ states and the higher-energy side $L_{HE2}$ to the transition between O $2p$ to Ni $3d_z^2$ states.

Figure 1b also reveals that both dip and peak structures red shift with decreasing temperature, by ~ 0.1 eV from 300 to 10 K (see also in the inset of Fig. 2c). The dip and peak positions in ΔR/R curves deviate slightly from the fitted $E_{HE1}$ and $E_{HE1}$ values, as the broad $L_{HE1}$ and $L_{HE2}$ curves overlap with each other.

## B. DENSITY-WAVE GAPS AND KINETICS

Figures 2a and 2b present temperature-dependent red shifts of $E_{HE1}$ and $E_{HE2}$, extracted from Fig. 1b. The sharp drops of both values around $T_{DW}$ strongly suggest that the energy red shifts originate from the DW order. Specifically, the DW order formation, accompanied by the opening of an energy gap Δ, gives rise to a band splitting of 2Δ[52-54]. Then the photoexcited carriers temporarily accumulate on the top band branch due to the bottleneck effect. Consequently, the measured high-energy electronic excitations shift by 2Δ from above to below $T_{DW}$.

Figure 2c depicts the temperature-dependent DW gaps Δ(T) calculated from Figs. 2a and 2b, by using the relation $E_{HE}(T) = E_{HE}(300\ K) - 2\Delta(T)$. The extracted Δ(T) below $T_{DW}$ can be well fitted by the BCS-type gap $\Delta(T) = \Delta(0) tanh(1.74\sqrt{T_c/T - 1})$, yielding two gaps $\Delta(0)_{HE1} = 54 \pm 2$ meV and $\Delta(0)_{HE2} = 67 \pm 3$ meV. Here, $\Delta(0)_{HE1}$ is consistent with optical conductivity [18], while $\Delta(0)_{HE2}$ aligns with previous ultrafast spectroscopy results[18,47].

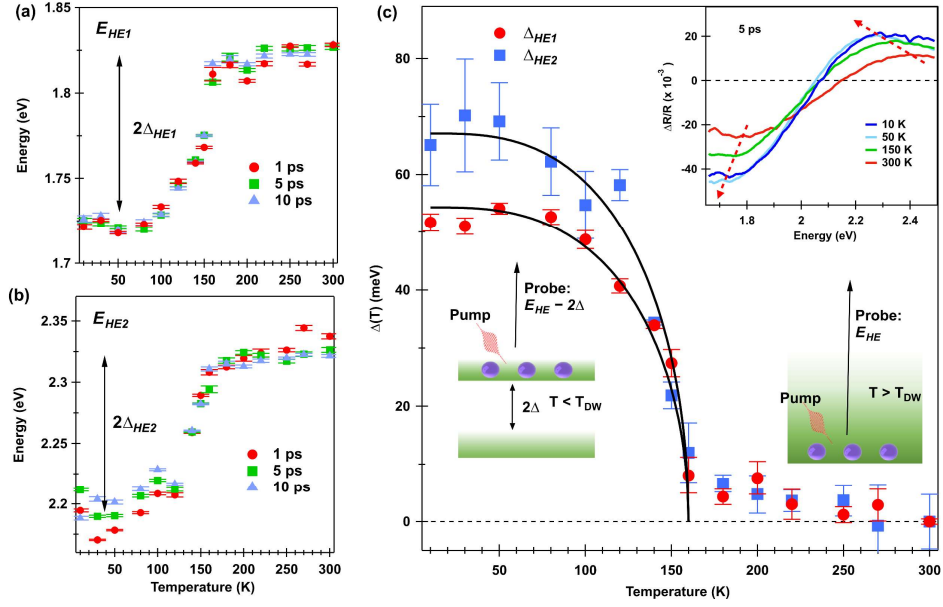

FIG. 2. Temperature evolution of high energy electronic excitations and DW gaps. (a) and (b) Temperature-dependent red shifts of $E_{HE1}$ and $E_{HE2}$ below the DW transition, obtained from the Lorentzian fits of horizontal cuts at delays of 1, 5 and 10 ps in Fig. 1b. Error bars are obtained from fits. (c) DW gaps obtained from the averaged red shifts in (a) and (b). $\Delta_{HE1}$ and $\Delta_{HE2}$ are DW gaps calculated by using the equation $E_{HE}(T) = E_{HE}(300 \text{ K}) - 2\Delta(T)$. The black curves are fits by the BCS-type gap. The right-upper inset indicates typical cuts at 5 ps (extracted from Fig. 1a), with energy red shifts marked by red arrows. The lower insets illustrate bands with (without) a splitting of $2\Delta$ below (above) the DW transition. With DW gap opening, the photoexcited hot electrons transiently accumulate on top of the band splitting, giving rise to $2\Delta$ red shifts.

Figure 3a and 3b display the temperature-dependent kinetics of *HE1* and *HE2*, respectively, consisting of superimposed oscillations and decays. The oscillations corresponding to coherent Raman-active phonon modes, will be discussed later. The decays are well fitted by a single-component exponential function and a linear background above the DW transition $\Delta R/R = A_f\exp(-t/\tau_f) + A + B \times t$, whereas a two-component exponential function $\Delta R/R = A_f\exp(-t/\tau_f) + A_s\exp(-t/\tau_s) + A + B \times t$, is necessarily required in DW state. Here $\tau_f$ and $\tau_s$ correspond to the fast and slow decays, and the linear background is included in the fits to account for the contribution of a slow acoustic phonon, which dominates the signal at longer timescales (see more details in Supplemental Material, Section 2). The temperature-dependent intensities and decay times for the slow decay components of *HE1* and *HE2*, which appear only in DW state, are shown in Figs. 3c and 3d, respectively. The sharp rise of the slow components below $T_{DW}$ indicates that these signals originate from the bottleneck effect following the opening of DW gaps, consistent with previous ultrafast results by using monochromatic/narrowband lights[47].

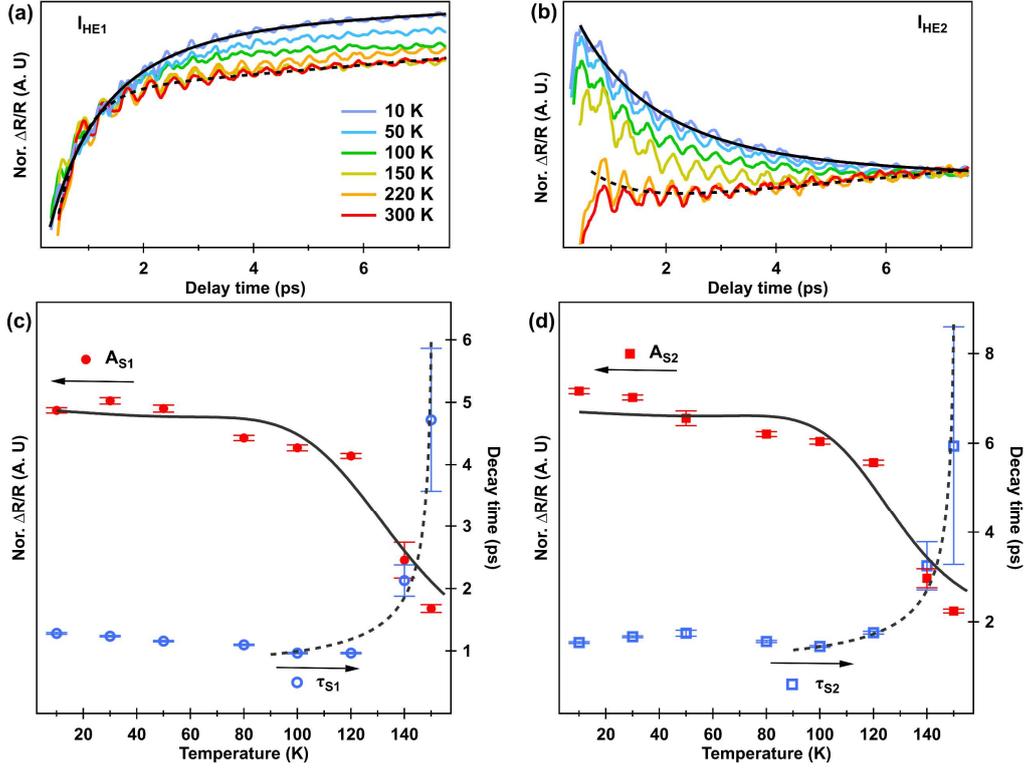

FIG. 3. Temperature-dependent kinetics of *HE1* and *HE2*. (a) and (b) Intensity of Lorentzian dips (*HE1*) and peaks (*HE2*) as a function of delay time, obtained from Fig. 1b by using Lorentzian fits. The black dashed lines are fits with a sub-ps exponential decay and a linear background, for the data above DW transition. The black solid curves are fits using a two-component exponential function and a linear background, for the results below DW transition. (c) and (d) Amplitude and decay time of the slow component, obtained from the fits in (a) and (b). Error bars are obtained from the fits. The black solid curves are fits by the RT model, producing $\Delta(0)_{HE1} = 53 \pm 1$ meV and $\Delta(0)_{HE2} = 64 \pm 1$ meV. The black dashed curves are fits for the decays approaching the DW transition temperature, by $\tau \propto 1/\Delta(T)$, where $\Delta(T)$ is the BCS-type gap.

The over-gap recovery process of bottlenecked electrons is dominated by bosons with energy higher than $2\Delta$. When the temperature approaches the DW transition, more low-energy bosons are involved due to the closing of gaps, giving rise to a divergence with decay time inversely proportional to the gap size, $\tau \propto 1/\Delta(T)$, as shown in Fig. 3c and 3d). Based on the RT model, the temperature dependence of the amplitude of the slow decay $A_{s1}$ and $A_{s2}$, which appear only in DW state, can be simulated as[28]

$$A(T) = \frac{\epsilon_I/(\Delta(T) + k_B T/2)}{1 + B\sqrt{\frac{2k_B T}{\pi \Delta(T)}} exp(-\Delta(T)/k_B T)}$$

where $\epsilon_I$ is the pump intensity per unit cell, $B$ is the parameter determined by boson numbers and cutoff frequency. Using the BCS-type gap $\Delta(T)$, the fits produce two gaps $\Delta(0)_{HE1}$ and $\Delta(0)_{HE2}$ as well (Fig. 3c and 3d), consistent with the results in Fig. 2c.

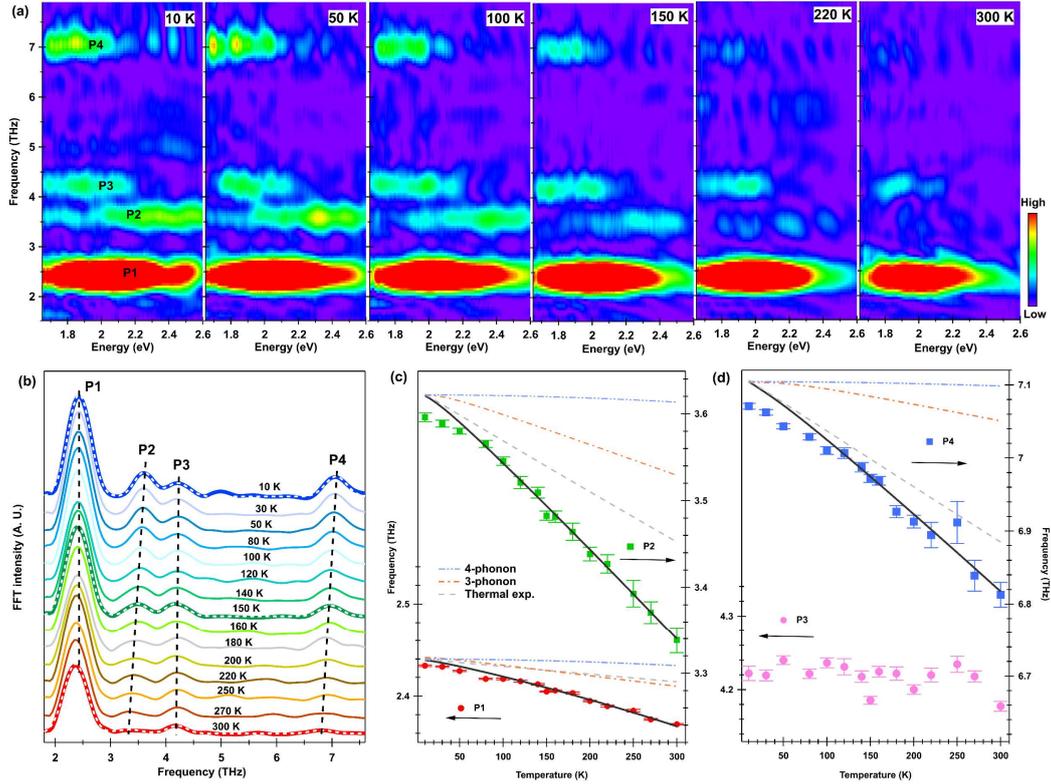

FIG. 4. Temperature-dependent coherent phonons. (a) FFT 2D images, calculated from ΔR/R results in Fig. 1b. Four peaks are marked as *P1*, *P2*, *P3* and *P4*. (b) Integrated FFT intensities over the range from 1.7 to 2.6 eV. The white dashed curves are Gaussian fits with four peaks. The black dashed curves serve as guides for the temperature evolution of phonon peaks. (c) and (d) Softening of coherent phonons with increasing temperature. The black solid lines are fits by Eq. 3 over the range from 100 K to room temperature, consisting of thermal expansion (the grey dashed lines), three-phonon (the red dashed lines) and four-phonon (the blue dashed lines) processes.

## C. TEMPERATURE-DEPENDENT COHERENT PHONON MODES

Figure 4a shows fast Fourier transform (FFT) analysis of the coherent oscillations at different temperatures. Four well-defined FFT peaks are identified at 10 K: *P1* = 2.43 THz (81 cm$^{-1}$), *P2* = 3.59 THz (120 cm$^{-1}$), *P3* = 4.22 THz (141 cm$^{-1}$), and *P4* = 7.07 THz (236 cm$^{-1}$), consistent with the Raman-active phonon modes from calculations and experiments [55,56]. These modes are characterized by in-phase and out-of-phase vibrations of the in-plane oxygen atoms, associated with the vibrations of nickel and apical oxygen atoms[55,56]. *P1* contributes to the oscillations in both *HE1* and *HE2*, while *P2* predominantly appears in *HE2*, and *P3* and *P4* appear only in *HE1*.

Figures 4c and 4d show temperature-dependent frequencies of all four phonons, obtained from the integrated FFT curves Fig. 4b. Upon increasing temperature, phonon softening appears for *P1*, *P2,* and *P4*, associated with kink-like features at ~100 K, while *P3* remains almost temperature-independent. Two main effects are considered to account for the softening of a Raman mode: thermal expansion and anharmonic phonon-phonon coupling. Thermal expansion, which originates from the volume or lattice changes of the material, can be described by the Grüneisen constant model, expressed as[57-59]

$$\Delta\omega_{TE} = (\frac{d\omega}{dV})_V \Delta V = \omega_0 exp(-n\gamma \int_0^T \alpha dT) - \omega_0 \quad (1)$$

where $\omega_0$ is the phonon frequency at zero temperature, $n$ is the degeneracy, $\gamma$ is the Grüneisen parameter, and $\alpha$ is the thermal expansion coefficient.

Anharmonic phonon-phonon coupling, originating from cubic and quartic anharmonic terms in the lattice potential, leads to anharmonic decays into two or three phonons, which are called three- and four-phonon processes, respectively. Within the Klemens model, this term can be expressed as[58-62],

$$\Delta\omega_A = (\frac{d\omega}{dT})_A \Delta T = d(1 + \frac{2}{e^x-1}) + f[1 + \frac{3}{e^y-1} + \frac{3}{(e^y-1)^2}] \quad (2)$$

where $x = \hbar\omega/2k_BT$, $y = \hbar\omega/3k_BT$, and $d$ and $f$ are contributions of three- and four-phonon processes, respectively.

Taking the above two effects into account, the temperature-dependent softening of a Raman mode can be expressed as [58,59,61]

$$\omega(T) = \omega_0 + \Delta\omega_{TE} + \Delta\omega_A \quad (3)$$

As shown in Figs. 4c and 4d, phonon softening of *P1*, *P2* and *P4* can be well fitted by Eq. 3 over the range from 100 K to room temperature, using constant $\gamma$ and $\alpha$ values. According to these fits, the four-phonon process is negligible since *f* is the approximately an order of magnitude lower than *d* (See fitted parameters in Supplemental Material, Section 3). Both thermal expansion and three-phonon process contribute almost equally to the softening of *P1*. For *P2* and *P4*, thermal expansion accounts for a shift of about 0.18 and 0.21 THz from 10 to 300 K, corresponding to about 65% and 75% of the total softening, respectively. Notably, deviations between the fits and measured phonon frequencies become evident at cryogenic temperatures below 100 K. At cryogenic temperatures, the phonon intensities of *P1*, *P2* and *P4*, also exhibit clear deviations from the linear extension of the high temperature results (See more details in Supplemental Material, Section 4). These observations point to an additional contribution, possibly from electron-phonon coupling, which becomes significant in DW state.

## III. DISCUSSION

The red shifts of *HE1* and *HE2* below the DW transition reveal the existence of two DW gaps in La$_3$Ni$_2$O$_7$, reminiscent of the two-gap structure in superconducting states under pressure revealed from Andreev reflection[63]. Notably, a recent theoretical study has proposed that an electric field can drive superconductivity above liquid nitrogen temperature in La$_3$Ni$_2$O$_7$ films at ambient pressure[64]. As a non-contact and bulk-sensitive experimental method, time-resolved optical spectroscopy with a WLC broadband probe is well-suited for direct probing the evolution of band-dependent dynamics of DW and superconducting gaps, especially when integrated with high-pressure or electric-gating setups.

The observed coherent phonon oscillations, which soften upon increasing temperature and persist even above the DW transition, are distinct from the DW collective modes (e.g. the amplitude mode in CDW materials), which only appear below the transition temperature[30]. The absence of strong DW collective modes suggests a novel DW mechanism in nickelate materials. At high temperatures, the phonon softening is likely induced by a combination of thermal expansion and phonon-phonon coupling, whereas at cryogenic temperatures, electron-phonon coupling should be taken into account. The energy-dependent behavior of the coherent oscillations reveals distinct coupling strength between different phonon modes and electronic bands.

## IV. CONCLUSIONS

In summary, we have observed two high-energy electronic excitations and band-dependent DW gaps in $La_3Ni_2O_7$ bulk crystal at ambient pressure from time-resolved optical spectroscopy with a WLC broadband probe. Four band-dependent coherent phonon oscillations are well resolved, and upon increasing temperature, three of them exhibit robust softening. Analysis of the phonon softening provides critical information on thermal expansion, phonon-phonon coupling and electron-phonon coupling. This study highlights the complex gap and phonon dynamics in nickelate materials, which are characterized by multiple electronic bands and many phonon properties.

## ACKNOWLEDGEMENTS

This work was supported by funding from the National Key Research and Development Program of China (Grant No. 2022YFA1204700), the National Natural Science Foundation of China (Grants No. 12250710126, 92056204 and 21902135), and the Quantum Science and Technology-National Science and Technology Major Project (2023ZD0300300).

## DATA AVAILABILITY

The data that support the findings of this article are not publicly available. The data are available from the authors upon reasonable request.

## Section 1. Time-resolved optical spectroscopy

In our home-built time-resolved optical spectroscopy, the ultrafast laser pulses are produced from a Ti:sapphire amplifier system (Coherent, Astrella, 1 kHz repetition rate, a center wavelength of 800 nm, and 35 fs). The output laser beam is split into two separate parts for pump and probe paths. The probe goes is focused into a sapphire crystal to generate a white-light-continuum (WLC) broadband light, ranging from 450 to 750 nm. Both pump and probe beams are onto the surface of the bulk $La_3Ni_2O_7$ crystal, which is installed in a cryostat. The pump beam is modulated by a chopper running at a frequency 500 Hz and synchronized to the laser amplifier. Finally, the reflected probe signals with both pump on and off ($I_{on}$ and $I_{off}$), are collected by a spectrometer, to obtain the transient reflection changes using $\Delta R/R = (I_{on} - I_{off})/I_{off}$.

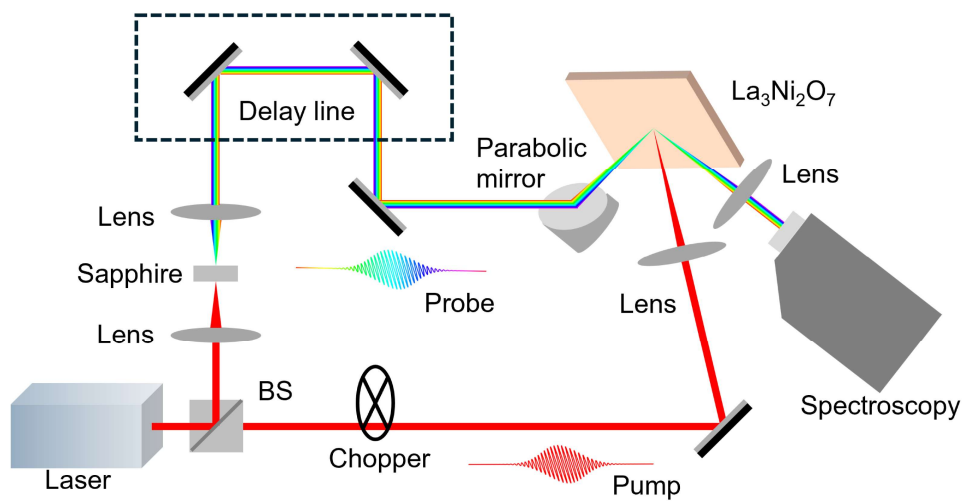

Figure S1. Illustration of time-resolved optical spectroscopy setup.

## Section 2. Long time dynamics and acoustic phonon

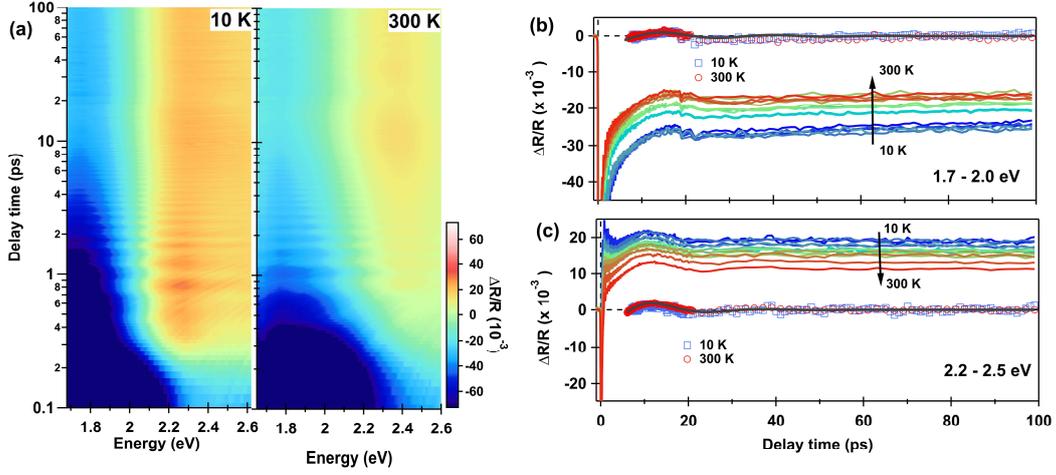

Figure S2. Time-resolved optical results up to 100 ps. (a) 2D map of ΔR/R at 10 and 300 K, with the same color scale. (b) and (c) Averaged ΔR/R versus delay time for negative (1.7 - 2.0 eV) and positive (2.2 - 2.5 eV) components, respectively. The red circles and blue squares are pure oscillatory parts by subtracting an exponential background. The black curves are fitted by a highly damped cosine function, with a frequency of ~40 GHz.

## Section 3. The obtained fitting parameters

Table 1. Parameters obtained from fits by Eq. 3.

| Phonon | $n\gamma\alpha$ | $d$ | $f$ |
|---|---|---|---|
| P1 | 3.4E-5 ± 1E-6 | 3.3E-3 ± 2E-4 | 9.1E-5 ± 1E-6 |
| P2 | 1.6E-4 ± 2E-5 | 1.5E-2 ± 1E-3 | 2.4E-4 ± 6E-5 |
| P4 | 1.0E-4 ± 1E-5 | 2.0E-2 ± 2E-3 | 6.5E-4 ± 5E-5 |

## Section 4. FFT intensities of coherent phonons

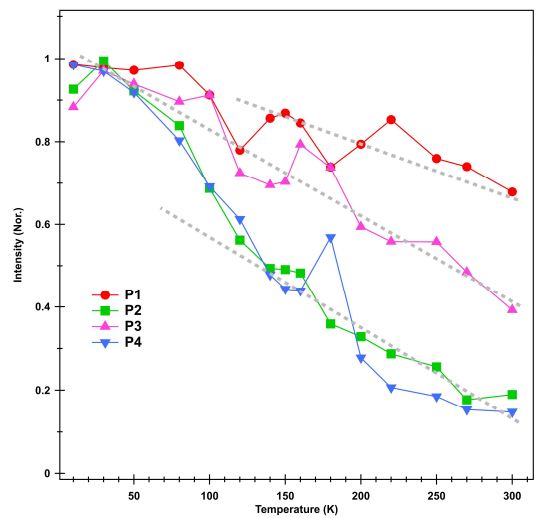

Figure S3. FFT intensity as a function of temperature for coherent phonons, extracted from Fig. 4b by Gaussian fits. The dashed grey lines follow the results at temperatures above the DW transition, and highlight the deviations below the DW transition.